\documentclass[twocolumn,aps,prl,groupedaddress,amssymb]{revtex4}
\usepackage{graphicx}
\usepackage{dcolumn}
\usepackage{amsmath}

\newcommand{\Ht}{H_{\rm t}}
\newcommand{\kB}{k_{\rm B}}
\newcommand{\VA}{\frac{V}{A}}
\newcommand{\VAt}{\left(\frac{V}{A}\right)_{\rm t}}
\draft
\begin{document}

\title{Instability of human societies as a result of conformity}

\author{A. L. Efros}
\email{efros@physics.utah.edu}
\affiliation{University of Utah, Salt Lake City UT, 84112 USA}

\author{ P. D\'{e}sesquelles}

\affiliation{Institut de Physique Nucl\'{e}aire, 15 rue Georges Cl\'{e}menceau,
F91406 Orsay France.}

\date{\today}

\begin{abstract}

We introduce a new model that mimics the strong and sudden effects induced by
conformity in tightly interacting human societies. Such effects range from mere
crowd phenomena to dramatic political turmoil. The model is a modified 
version  of the Ising Hamiltonian. We have studied the properties of this
Hamiltonian using both a Metropolis simulation and analytical derivations. Our
study shows that increasing the value of the conformity parameter, results in a
first order phase transition. As a result a majority of people begin honestly
to support the idea that may contradict the moral principles of a normal human
beings though each individual would support  the moral principle without tight
interaction with the society. Thus, above some critical level of conformity our
society occurs to be instable with respect to   ideas that might be  doubtful.
Our model includes, in a simplified way, human diversity with respect to
loyalty to the  moral principles.\\


\end{abstract}

\maketitle

\section{Introduction}

It is commonly observed that strongly interacting groups of people may show
very abrupt and unexpected behavior or opinion changes. These "transitions" may
even lead people to behave against their inmost sense of good and evil. This is
seen in many crowd phenomena, less harmfully, in fashion phenomena, or, more
dramatically, in the setting off of civil wars often leading to despotic
regimes. Of course,  these phenomena may have very different reasons, but we
propose a very simple model that shows the same features and which may give
some insights on the mechanisms of these regime reversals. Our hope is based
upon similarities in the manifestations of different non-democratic regimes
which suggest that similar universal mechanisms may have played a role.

All those regimes had powerful and merciless mechanisms of repression against
people who were either enemies of the regime or merely neutral to it. However,
one could argue that (see, for example, \cite{king}) the stability of those
regimes was not based only on repression. In some cases, the honest belief of
the majority of the population in the official ideas, or, at least the fact
that people  behave as if it was the case, could play a major role. For
example, as soon as this faithfulness became weak, the Soviet Union fell and
the security forces were unable to stop this process. In many cases these
ideologies drastically contradict the moral principles of a normal human being.
It is reasonable to think that those principles are shared by a vast majority
of human beings. The experience of the 20th century thus shows that under
certain conditions these principles fail and the behavior of people seems to be
controlled by some collective phenomena. Once more, we do not pretend that the
model we have developed \emph{explains} these historical events. Our modest
goal is to show its \emph{analogies} with human situations and to show how
conformity can lead to unexpected extreme and paradoxical situations.

These sociological or historical phenomena are often sudden, turmoil is
drastic, and its apparent cause seems insignificant. For a physicist, this
problem resembles a phase transition in a system consisting of interacting
elements. In the system under study these elements are people and the most
important part of their interaction is conformity. Conformity implies the
ability, or the propensity, of people to change their opinion under influence
of the opinions of surrounding people. The  conformity may be  measured
quantitatively by a simple experiment of the following type. A few pieces of
paper each with its own number are demonstrated to a group of people. Each
piece is either white or black. After the pieces are removed the
experimentalist asks what is the color of piece \#2. In fact, only one person
in a group is being tested in this experiment; the rest of the group are also
experimentalists, though the tested person is unaware of that. The tested
person is asked the question after all the experimentalist's team presents the
same wrong answer one after another. This experiment shows how difficult  is to
say that white is white after all other members of a group claimed that the
white piece is black. The majority of people are unable to remain true to their
conviction and will give a wrong answer (when the tested persons are alone,
they almost always give the right answer). As far as we know, these studies
began in the  middle 50th. Nowadays  they flourish mostly in the purpose of
marketing. Now  everybody can take  a free test to estimate roughly  his/her
level of conformity using  Asch's Conformity Study \cite{w2}.  The role of
conformity in a society has been considered previously \cite{callen,hogan}.

We do not wish to imply that conformity itself is something bad. This is one of
human properties which allows social life. However, we study below how
conformity may lead to instabilities in human societies.

\section{ Statistical approach}

Statistical physics is widely used for studying opinion dynamics in society in
the framework of different models\cite{mob,mol,lat}. There are good reviews of
the field\cite{di,ho}. Our paper is devoted to a different aspect of the
problem, and we use here rather the Gibbs ensemble than dynamical equations.  

Our rough model considers the opinion of the society with respect to only one
statement. We assume that all previous experience of the society including the
inherited ideas of good and bad dictates a negative opinion with respect to
that statement. However there is a conformity interaction between overlapping
groups of the society and also there are some individuals who, in spite of
everything, have a positive opinion with respect to the statement. Their
positive opinion is a little bit stronger than the negative opinion of other
people. Following the terminology of Mobilia \cite{mob} we shall call these
individuals ``zealots''. The number of zealots is not fixed and is determined
from  the thermal equilibrium. We interpret the phase transition in this system
as social instability. At small conformity the majority has a negative opinion
while at high conformity the opinion of the majority becomes strongly positive,
i.e. the majority of people become zealots.

The Hamiltonian of the model reads:

\begin{equation}
 H=\sum_i \phi_i\,q_i- \frac{1}{2}\,\sum_{i\neq j}V_{i,j}\,q_iq_j\,.
\label{H}
\end{equation}

For simplicity it is formulated on a two-dimensional lattice. Variables $q_i$
may take values $-1/2$ or $1/2+\alpha$. They represent the opinion of the 
person on a site $i$ with respect to a given statement. In terms of the
previous terminology zealots have positive opinions and value $\alpha>0$. This
parameter describes some extra strength of the positive opinion, the negative
opinion being the "normal" one (whether because it corresponds to the true
color of the piece of paper, or to the opinion dictated by the "universal"
notion of good and evil\ldots). In the absence of conformity ($V=0$), at low
temperatures all the opinions are negative (within thermal fluctuations).

The random values $\phi_i$ are {\em positive} with a Gaussian distribution
$F(\phi)=\sqrt{2}/(\sqrt{\pi}A) \exp {-(\phi^2/2A^2)}$. They describe the
diversity of the individuals with respect to the statement under consideration.
Indeed, considering only the first term in Eq. (\ref{H}), one finds that the
energy $H$ has a minimum when all values of $q_i$ are negative. The second term
describes interaction and conformity. It has a minimum when all $q_i$ have the
same sign, in other words when all people think in the same way. Moreover, at
any positive value of $\alpha$, it prefers all people to be zealots. For
simplicity we consider only the interactions with the four nearest neighbors on
a square lattice and $V_{i,j}=V$.

Eq. (\ref{H}) can be represented as an Hamiltonian of the Ising model in some
fictitious magnetic field\cite{di1}. Introducing the new variables
$S_i=q_i-\alpha/2$ one gets:

\begin{equation}
H=\sum_i S_i h_i- \frac{1}{2}\sum_{i\neq j}V_{i,j}S_i S_j +C,
\label{dit}
\end{equation}

where:

\begin{equation}
h_i=\phi_i-\sum_jV_{i,j}\alpha/2
\label{h}
\end{equation}

is a random ``magnetic field'' at a site $i$, C is is irrelevant constant,
and the values of $S_i$ are $\pm 1/2(1+\alpha)$.

\section{First order phase transition}

We now show that our system has a first order phase transition. At zero
temperature it can be derived exactly. To find the transition point we compare
the energies of the two states. In the state I all $q_i=-1/2$. It has the
energy per site:

\begin{equation}
E_I=-{\frac{1}{2}}\,\langle \phi \rangle-\frac{Vz}{8}\,. \label{1}
\end{equation}

In the state II all $q_i=1/2 +\alpha$ which means all people are zealots. It
has an energy per site:

\begin{equation}
E_{II}=\left({\frac{1}{2}}+\alpha\right)\,\langle \phi \rangle-\frac{Vz}{2}\,
\left({\frac{1}{2}}+\alpha\right)^2\,. \label{2}
\end{equation}

Here $\langle \phi \rangle=\int_0^\infty \phi\,F(\phi)\,d\phi$ and $z=4$ is the
number of nearest neighbors taking part in the interaction with a given site.
When calculating the interaction per site we must take $V/2$ since this
contribution belongs to two sites. It is easily found that:

\begin{equation}
\langle \phi \rangle=\sqrt{\frac{2}{\pi}}\,A\,. \label{p}
\end{equation}

One can see that, at small interaction $\alpha\, z\, \VA\ll 1$, the state I has
the lower energy while in the opposite case the state II has the lower energy
and represents the ground state of the system. This is a typical situation for
a first order phase transition. It happens at $E_I=E_{II}$. Using Eqs.
(\ref{1},\ref{2},\ref{p}) one finds that the transition occurs at:

\begin{equation}
\label{at} \VAt\approx \frac{1.6}{\alpha\, z}
\end{equation}

and the energy in the transition point is given by the equation:

\begin{equation}
\Ht=-\frac{NVz}{8}\,\left(1+2\alpha\right)\,, \label{ht}
\end{equation}

where $N$ is the number of sites in the system. Note that this result can be
obtained from the condition $< h> =0$, where $h$ is given by  Eq.(\ref{h}).

If conformity is small $(\VA<\VAt$), the total energy can be found from Eqs.
(\ref{1},\ref{p}). It has the form:

\begin{equation}\label{small}
H_{I}=-\frac{1}{2}V N\left[
\sqrt{\frac{2}{\pi}}\left(\VA\right)^{-1}+\frac{z}{4}\right]\,.
\end{equation}

The total energy at large conformity can be found from Eqs. (\ref{2},\ref{p}).
One obtains:

\begin{equation}
\label{large} H_{II}=V N\,\left(\frac{1}{2}+\alpha\right)\left[
\sqrt{\frac{2}{\pi}}\,\left(\VA\right)^{-1}-\left(\frac{1}{2}+\alpha\right)\,
\frac{z}{2}
\right]\,.
\end{equation}

\section{Monte-Carlo simulation at finite temperature}

A Metropolis \cite{met} code has been written for the simulation of the
Hamiltonian (Eq. (\ref{H})) on a 2D ($L\times L=N$) square periodical lattice.
At each Monte-Carlo step, a site $k$ of the lattice is chosen at random. The
value of $q_k$ of the site is flipped with a probability
$P_k=\operatorname{Min} (1,\exp{-\delta H_k/\kB T})$ where $\delta H_k =
\epsilon_k\, \delta q_k$ is the  energy cost of flipping. Here:

\begin{equation}
\label{eps}
 \epsilon_k=\phi_k-\sum_j V_{kj}\ q_j.
\end{equation}

It is well known that, after an initial number of Monte-Carlo steps that are
necessary for relaxation, the averaging of any function of the $q_i$ over the
subsequent set of distribution of $q_i$, obtained by this way, is equivalent to
averaging over the Gibbs ensemble. Depending on the conditions, the operation
is performed from one hundred to one million times per site. The simulation has
been run for different values of the independent parameters $T,\ \VA,\ \alpha$,
starting from different initial distributions of the $q_i$. In what follows we
use, instead of $T$, a dimensionless temperature, so that $T\rightarrow k_B
T/V$.

\begin{figure}[htbp]
\begin{center}
\includegraphics[width=7cm]{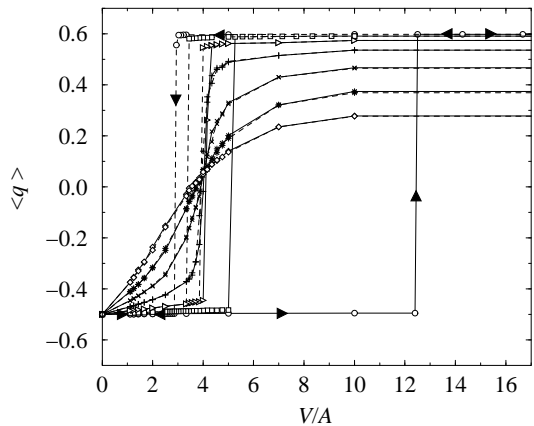}
\caption{\small Average public opinion as a function of $\VA$ at different
temperatures (circles: 0.4, squares: 0.5, triangles: 0.6, plus: 0.7,
crosses: 0.8, stars: 0.9, diamonds: 1.0) as given by the simulation. The
results obtained for increasing conformity are connected by a solid line, those
corresponding to the backward path are connected by dotted lines. For
temperatures greater than 0.8 both paths are superimposed (within numerical
fluctuations).}\label{Fig q(V/A)}
\end{center}
\end{figure}

The main results of the simulation are shown in Fig. \ref{Fig q(V/A)}. The time
averaged value $\langle q \rangle$ is plotted against $\VA$ at various values
of the temperature at $\alpha=0.1$. One can see that, at low temperatures and
low conformity the point of view of the population with respect to the
statement under study is negative. At strong conformity ($\VA\rightarrow
\infty$), all members of the population become zealots. At low temperatures one
can see a wide hysteresis. At higher temperatures it disappears and all curves
intersect at one point at $\VA \approx 4$. This value is in a good agreement
with Eq. (\ref{at}) at $z=4$ and $\alpha=0.1$. The value of $\langle q \rangle$
in the transition point is $\approx 0.05$ (i.e. $(-1/2+1/2+\alpha)/2$). The
hysteresis loop collapses with increasing temperature.

\begin{figure}[htbp]
\begin{center}
\includegraphics[width=7cm]{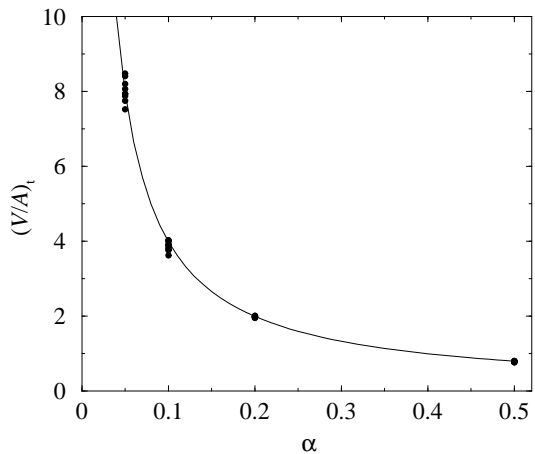}
\caption{\small Transition value $\VAt$ as a function of $\alpha$. The line
corresponds to Eq. (\ref{at}) and the dots to the simulation results obtained
for different values of the temperature ($T\in[0,1]$).} \label{Fig At(alpha)}
\end{center}
\end{figure}

Fig. \ref{Fig At(alpha)} shows the simulation results for $\VAt$ as a function
of $\alpha$ at different temperatures. One can see that, at the transition, it
reproduces very well Eq. (\ref{at}) which was obtained at zero temperature (the
same result can be obtained writing that at the transition $\langle \epsilon
\rangle = 0$, Eq. (\ref{eps})). The simulation shows that Eq. (\ref{at}) is
applicable over a wide temperature range. This follows also from the fact that
curves corresponding to different temperature cross at one point in Fig.
\ref{Fig q(V/A)}.

\begin{figure}[htbp]
\begin{center}
\includegraphics[width=7cm]{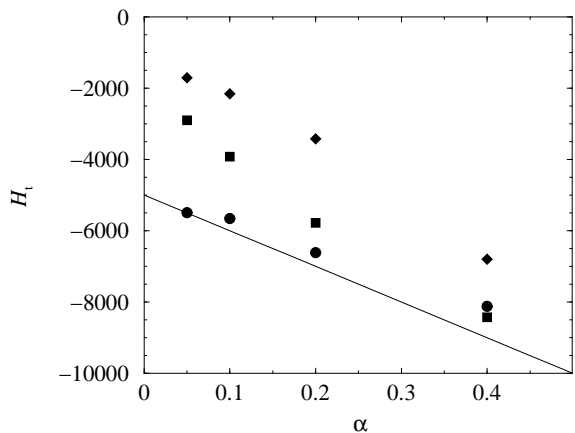}
\caption{\small Transition value $\Ht$ as a function of $\alpha$. The line
corresponds to Eq. (\ref{ht}) derived at $T=0$. The points are obtained from
the simulation for different values of the temperature (diamonds:  $T=1$,
squares: $T=0.7$, circles: $T=0.1$).} \label{Fig Ht(alpha)}
\end{center}
\end{figure}

However the energy at the transition point differs substantially from the
zero-temperature energy given by Eq. (\ref{ht}). The transition energy $\Ht$ is
shown in Fig. \ref{Fig Ht(alpha)}. To avoid the auto-magnetisation effect at
low temperature, the initial directions of the spins are sorted at random.

\begin{figure}[htbp]
\begin{center}
\includegraphics[width=7cm]{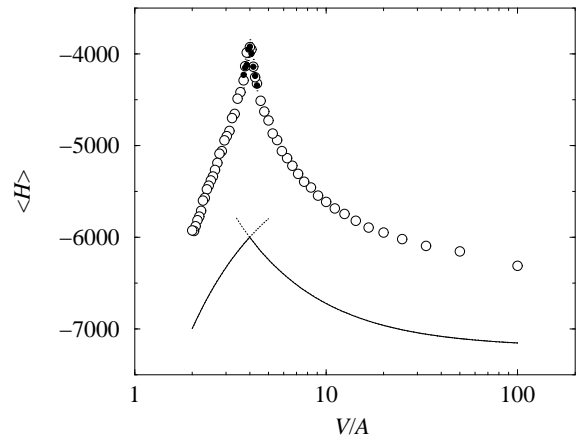}
\caption{\small Total energy as a function of $\VA$ in the case $T=0.7,\
\alpha=0.1$ obtained on a $100\times 100$ lattice (open dots). The filled dots
correspond to one quarter of the energy obtained for a $200\times 200$ lattice.
The zero-temperature results given by Eqs. (\ref{small},\ref{large}) are shown
by a solid line.}\label{Fig H(A)}
\end{center}
\end{figure}

It is also interesting to compare the behavior of the energy as a function of
$\VA$ at a finite temperature with the zero-temperature behavior as given by
Eqs. (\ref{small},\ref{large}). These results are given in Fig. \ref{Fig H(A)}.

\section{Discussion of computational results}

As far as we know, the Hamiltonian given by Eq. (\ref{H}) has never been
studied before. At $A=\alpha=0$ it coincides with the ferromagnetic Ising
Hamiltonian and it has a second-order phase transition in the two-dimensional
case. We claim that our Hamiltonian has a first order transition at a point
given by Eq. (\ref{at}). It is obvious at $T=0$. Convincing argument is a
singularity in the behavior of energy as a function of $\VA$ shown in Fig.
\ref{Fig H(A)} that we interpret as a discontinuity of the first derivative of
the energy in the transition point. Our understanding of the hysteresis is that
our modeling in real time goes very slowly and therefore we may see a
continuation of the phase from which we start above the transition point. That
is typical for the first order transition and reminds super-cooled liquid.

On the other hand, we study a very low value of $\alpha$ and transition occurs
at small value of $A$. Therefore one should expect that it should be close to
the second order phase transition in the Ising model. Therefore we may observe
clusters of zealots near the phase transition. As an example the distribution
of the zealots at the threshold of the transition is shown in Fig. \ref{Fig
spins}. One can see that  the system has a large correlation radius which is
typical for the second order phase transition. Thus, we think that we have a
weak first order phase transition.

\begin{figure}[htbp]
\begin{center}
\includegraphics[width=7cm]{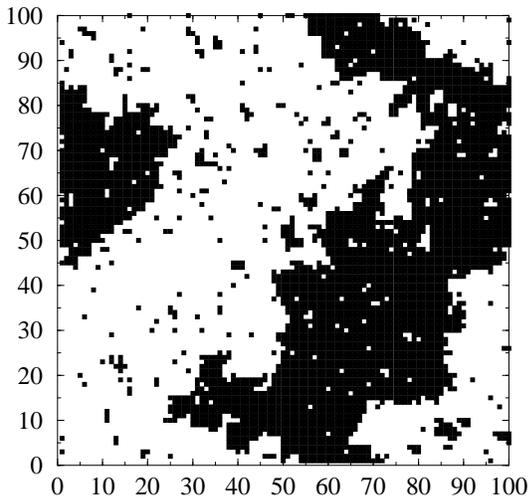}
\caption{\small Distribution of the zealots at $T=0.62,\ \VA=\VAt,\ N=10^4,\
\alpha = 0.1$. The final distribution of the zealots are indicated by black
squares (the 100$\times$100 lattice is periodical).}\label{Fig spins}
\end{center}
\end{figure}

\section{Conclusions}

We propose a simple Hamiltonian that models the drastic opinion changes that
can be experienced by human groups submitted to strong mutual influence, even
changes that contradict a cultural achievements of the past. This model shows
how, due to the  conformity, the group may express an opinion opposite to the
opinion which each of its isolated members would have. The model contains the
following parameters: $T,A,V,\alpha, z$ that can be determined by sociological
methods. The level of conformity may be checked by experiments similar to those
described in the Introduction. The ``temperature" of the society could be found
by studying time fluctuations of the public opinion.

Our model is oversimplified, and the Eq. (\ref{at}) for the critical value at
the first order transition point may be not accurate. Nevertheless, if the
mechanism of collective phenomenon in the human society can basically be
described our  model, the ratio $\VAt$ as given by Eq. (\ref{at}) might
be an important characteristic of group phenomena.

\end{document}